\documentstyle[aps,eqsecnum,prb,psfig,floats]{revtex}
\begin{document}
\draft
\twocolumn[\hsize\textwidth\columnwidth\hsize\csname@twocolumnfalse\endcsname

\title{Magnon dispersion and thermodynamics in $\mathrm{\mathbf{CsNiF_3}}$}

\author{J. Karadamoglou and N. Papanicolaou }
\address{Department of Physics, University of Crete, and 
Research Center of Crete, Heraklion, Greece}

\author{X. Wang\cite{XWAdd} and X. Zotos}
\address{Institut Romand de Recherche Num\'erique en
Physique des Mat\'eriaux (IRRMA), PPH-Ecublens, CH-1015 Lausanne, Switzerland}

\maketitle

\begin{abstract}
\hspace{5pt}
We present an accurate transfer matrix renormalization 
group calculation of the thermodynamics in a quantum
spin-1 planar ferromagnetic chain. We also calculate 
the field dependence of the magnon gap and confirm the accuracy 
of the magnon dispersion derived 
earlier through an $1/n$ expansion.
We are thus able to examine the validity of a number 
of previous calculations and further analyze a wide range 
of experiments on $\mathrm{CsNiF_3}$ concerning the magnon
dispersion, magnetization, susceptibility, and specific
heat.
Although it is not possible to account for all data 
with a single set of parameters, the overall qualitative 
agreement is good and the remaining discrepancies may 
reflect departure from ideal quasi-one-dimensional 
model behavior.
Finally, we present some indirect evidence to the effect 
that the popular interpretation of the excess specific heat
in terms of sine-Gordon solitons may not be appropriate.
\end{abstract}
\vskip 1cm
]
\section{Introduction}
\label{sec:1}
The magnetic compound $\mathrm{CsNiF_3}$ undergoes three-dimensional
(3D) ordering at very low temperatures, $T < T_N=2.7 $ K,
but exhibits essentially one-dimensional (1D) behavior
for $T>T_N$. A number of experimental investigations \cite{1}
suggest that an appropriate 1D model is described by the spin 
$s=1$ Hamiltonian
\begin{equation}
\label{eq:1}
W = 
\sum_n \left[
	-J {\bf S}_n \! \cdot {\bf S}_{n+1}
	+ A ( {\mathrm{S}}_n^z )^2 
	- g \mu_B {\bf H} \! \cdot \! {\bf S}_n 
\right],
\end{equation}
which contains a ferromagnetic ($J>0$) isotropic exchange
interaction and an easy-plane ($A>0$) single-ion anisotropy,
in addition to the usual Zeeman term produced by an applied 
field ${\bf H}$.

The derivation of accurate theoretical predictions based 
on Hamiltonian (\ref{eq:1}) turned out to be more 
difficult than anticipated thanks to the strong quantum 
fluctuations that occur in this quasi-1D system.
In particular, the leading-order magnon dispersion 
derived within the usual $1/s$ expansion is too crude an 
approximation for $s=1$.
As a result, inelastic neutron scattering experiments
were analyzed \cite{2} mostly on the basis of an alternative
dispersion derived by Lindgard and Kowalska \cite{3} using 
a self-consistent approach that is designed to properly
account for single-ion anisotropy. Similarly, a large 
body of experimental data became available for thermodynamic 
quantities  such as magnetization, susceptibility, and 
specific heat, but a corresponding theoretical calculation
proceeded slowly.
To the best of our knowledge, the most accurate calculation
of thermodynamics was provided by Delica et al. \cite{4} 
based on a quantum transfer matrix, while comparable success was 
claimed more recently by Cuccoli et al. \cite{5} through 
a sophisticated semiclassical approach.
The above two papers also contain an extensive list of 
references to earlier work.

It is the aim of the present paper to derive theoretical 
predictions that are accurate to within line thickness 
and thus provide a safe basis for the discussion of 
various issues that have been raised during the long 
history of this subject.

In Sec. \ref{sec:2}, experimental data on the 
magnon dispersion are analyzed in terms of an unconventional
$1/n$ expansion \cite{6} which is shown to contain the 
Lindgard-Kowalska dispersion as a special case. The accuracy 
of the leading $1/n$ approximation is confirmed by an independent
calculation of the field dependence of the magnon gap
using a density matrix renormalization group (DMRG) method,\cite{7} 
while a discussion of anharmonic corrections within 
the conventional $1/s$ expansion is also included for comparison.
Thermodynamic quantities are calculated in Sec. \ref{sec:3}
by a powerful transfer matrix renormalization group (TMRG) 
algorithm \cite{8,9,10} which addresses directly the 
infinite-chain limit.
We are thus in a position to appreciate the relative accuracy 
of earlier calculations, analyze all available data and anticipate
results of possible future experiments, as well as challenge 
popular interpretations in terms of sine-Gordon solitons.
A brief summary of the main conclusions is given in 
Sec. \ref{sec:4}.

\section{The magnon dispersion}
\label{sec:2}
The standard spinwave theory is a method for calculating 
quantum corrections around the classical minimum of 
Hamiltonian (\ref{eq:1}) by a systematic $1/s$ expansion.
The $1/n$ expansion developed in Ref. \onlinecite{6} is of a 
similar nature, except that the corresponding ``classical'' 
minimum is a variational Hartree-like ground state that 
is more sensitive to the nature of single-ion anisotropy
and thus provides a more sensible starting point.
Hence one obtains an accurate magnon dispersion even if 
the $1/n$ series is restricted to the harmonic approximation.

For a field applied in a direction perpendicular to the $c$-axis,
e.g., ${\bf H} =(H,0,0)$, the magnon energy at crystal momentum 
$q$ is given by
\begin{equation}
\label{eq:2}
\omega_q = 
2 J \left\{
	(1+\varepsilon) (\frac{\alpha}{4\varepsilon}-\cos q)
	\left[
		\frac{\alpha}{4\varepsilon}(1+\varepsilon)
		- (1-\varepsilon) \cos q
	\right]
\right\}^{1/2}
\end{equation}
Here and in the rest of the paper we employ rationalized 
parameters for anisotropy and field,
\begin{equation}
\label{eq:3}
\alpha = A/J,
\qquad
h = g_{\perp} \mu_B H/J,
\end{equation}
while energy and temperature may be measured 
in units of the exchange constant $J$. The notation 
employed for the gyromagnetic ratio $g_{\perp}$ implies 
that the corresponding ratio $g_{\parallel}$ for a 
field parallel to the $c$-axis may be different.
Finally, the dimensionless parameter $\varepsilon$ 
in Eq. (\ref{eq:2}) is determined in terms of $\alpha$ 
and $h$ by the algebraic equation
\begin{equation}
\label{eq:4}
\varepsilon = 
\frac{\alpha(1-\varepsilon^2)^{1/2}}{2h +4(1-\varepsilon^2)^{1/2}}.
\end{equation}
One should add that derivation of systematic $1/n$ corrections
to the harmonic approximation (\ref{eq:2}) is possible \cite{6}
but unnecessary in the parameter range of current interest:
$\alpha, h< 0.5$.

At zero field, the root of Eq. (\ref{eq:4}) is 
$\varepsilon = \alpha/4$ which is inserted in Eq. (\ref{eq:2})
to provide a completely explicit expression for the magnon
dispersion. 
For nonzero field, Eq. (\ref{eq:4}) may be solved by simple 
iteration starting with $\varepsilon = 0$. In fact, the result 
of a single iteration,
\begin{equation}
\label{eq:5}
\varepsilon 
\simeq 
\frac{\alpha}{2h+4},
\end{equation}
is practically indistinguishable from the exact root of 
Eq. (\ref{eq:4}) for parameters such that $\alpha, h < 0.5$. 
The last remark becomes especially important if one notes 
that the dispersion obtained by inserting the approximate root 
(\ref{eq:5}) in Eq. (\ref{eq:2}) is precisely the magnon dispersion
derived earlier by Lindgard and Kowalska,\cite{3} applied for 
$s=1$, which was in turn employed for the analysis of experimental 
data from inelastic neutron scattering.\cite{2}

\begin{figure}
\centerline{\hbox{\psfig{figure=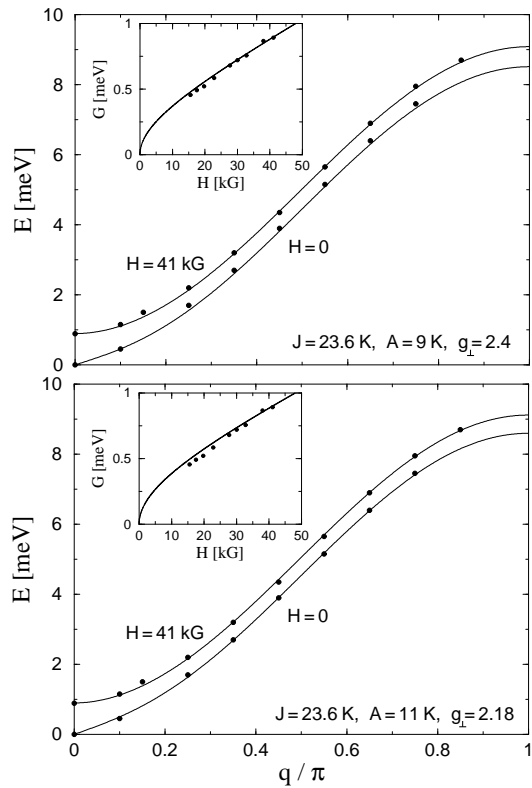,width=7.5cm}}}
\caption{
The magnon energy $E=\omega_q$ as a function
of crystal momentum $q$ calculated from Eq. (\protect\ref{eq:2})
for two values of the applied field, $H=0$ and $H=41 $ kG, 
and two different sets of parameters. The insets illustrate the 
corresponding field dependence of the $q=0$ magnon gap $G$ 
calculated from Eq. (\protect\ref{eq:8}). Solid circles 
represent experimental data from Ref. \protect\onlinecite{2} taken 
at $T=4.2 $ K.
}
\label{fig:1}
\end{figure}

The latter analysis provided what is often referred to as the 
standard set of parameters for $\mathrm{CsNiF_3}$:
\begin{equation}
\label{eq:6}
J=23.6 \ \mathrm{K},
\qquad
A=9 \ \mathrm{K},
\qquad
g_{\perp} = 2.4.
\end{equation}
The corresponding theoretical predictions of the magnon dispersion
(\ref{eq:2}) are compared to experimental data \cite{2} in the 
upper panel of Fig. \ref{fig:1}. 
The agreement is obviously very good for field $H=41 $ kG, 
while a slight but systematic deviation
is observed for $H=0$. This conclusion is somewhat surprising
in view of the claim in Ref. \onlinecite{2} that nearly perfect agreement 
is obtained for both field values, even though the Lindgard-Kowalska 
dispersion employed in the above reference is practically
identical to Eq. (\ref{eq:2}) for the set of parameters (\ref{eq:6}).
The systematic nature of this discrepancy makes it unlikely that 
the data communicated to us by Steiner \cite{11} differs from 
the data actually used in the analysis of Ref. \onlinecite{2}.
A more likely explanation is that the Lindgard-Kowalska dispersion
was further approximated by the authors of Ref. \onlinecite{2}, as is evident
in the expression for the magnon gap given in their Eq. (5).

Although the observed discrepancy appears to be minor, 
it nonetheless leads to a substantial redefinition of 
parameters.
Thus we have redetermined the exchange constant $J$ and 
anisotropy $A$ by a least-square fit of the zero-field 
data to dispersion (\ref{eq:2}), while the gyromagnetic ratio
was subsequently obtained by a one-parameter least-square 
fit of the $H=41 $ kG data.
The resulting new set of parameters 
\begin{equation}
\label{eq:7}
J=23.6 \ \mathrm{K},
\qquad
A=11 \ \mathrm{K},
\qquad
g_{\perp} = 2.18
\end{equation}
restores agreement with experiment for both field values, 
as is shown in the lower panel of Fig. \ref{fig:1}.
A notable feature of Eqs. (\ref{eq:6}) and (\ref{eq:7}) is 
that the exchange constant has remained unchanged. 
Indeed, throughout our analysis, we found no evidence for departure
of the exchange constant from the value $J=23.6 $ K which will thus be
adopted in the following without further questioning.

In contrast, the observed significant fluctuations in the 
anisotropy constant $A$ and gyromagnetic ratio $g_{\perp}$ 
simply reflect the fact that the magnon dispersion is not 
especially sensitive to those parameters. Therefore, their 
values given in either Eq. (\ref{eq:6}) or (\ref{eq:7}) cannot
be considered as established without further corroboration.
Now, the reduced value of the gyromagnetic 
ratio given in Eq. (\ref{eq:7}) is consistent with 
$g_{\perp}=2.1 \pm 0.05$ obtained independently by measuring 
the saturation magnetization at strong fields \cite{4} and is also
supported by the analysis of the zero-field susceptibility in Sec. 
\ref{sec:3}. But a proper choice of the anisotropy constant $A$
will be a matter of debate throughout this paper. In this respect,
one should keep in mind that the neutron data displayed in Fig.
\ref{fig:1} were taken at helium temperature, $T=4.2 $ K, which 
is relatively high but not too distant from the 3D-ordering
transition temperature $T_N=2.7 $ K. Hence, finite-temperature 
effects as well as deviations from ideal 1D behavior may 
already be present.

An important special case of the magnon dispersion (\ref{eq:2})
is the zero-momentum gap $G=\omega_{q=0}$, or 
\begin{equation}
\label{eq:8}
G = \left\{
	g_{\perp} \mu_B H \left[
		g_{\perp} \mu_B H + A \left(
			\frac{1+\varepsilon}{1-\varepsilon}
		\right)^{1/2}
	\right]
\right\}^{1/2},
\end{equation}
where we have made use of the algebraic equation (\ref{eq:3})
to simplify the expression.\cite{6} A comparison of the predictions 
of Eq. (\ref{eq:8}) with the measured field dependence of the magnon 
gap \cite{2,11} is shown in the insets of Fig. \ref{fig:1} for both
sets of parameters. Although the overall agreement is reasonable,
systematic deviations are present at relatively low 
field values in both cases. An attempt to redetermine the 
parameters by a least-square fit of the $q=0$ data 
to Eq. (\ref{eq:8}) yields values for $A$ and $g_{\perp}$ that 
would significantly compromise the agreement obtained at nonzero
crystal momentum $q$. 

Implicit in the preceding discussion is the presumption that 
the magnon dispersion (\ref{eq:2}) and its special case (\ref{eq:8})
are sufficiently accurate and there is no need to proceed with the
calculation of anharmonic $1/n$ corrections. We now test this 
assumption by a completely independent calculation of the field 
dependence of the magnon gap based on a density matrix 
renormalization group (DMRG) algorithm.\cite{7} An early effort
\cite{12} to apply a renormalization-group technique was restricted
to short chains (16 sites) and thus provided reasonable but not 
especially accurate estimates of the magnon gap. The DMRG algorithm
allowed us to calculate the gap on long chains up to $400$ sites.
We have also tested the stability of our results through Shanks or 
Richardson extrapolation \cite{13} and believe to have calculated 
the gap to an accuracy greater than the three figures actually 
displayed in the third column of Table \ref{tab:1}.
\begin{table}
\caption{
Magnon gap in units of $J$, for a typical anisotropy
$\alpha=A/J=0.38$, and a field $h = g_{\perp} \mu_B H / J$
applied in a direction perpendicular to the $c$-axis.
}
\begin{tabular}{cccc}
   $h$   & \multicolumn{3}{c}{Magnon gap  $G$ } \\
         &  $1/n$  & DMRG    &  $1/s$  \\
\hline
  0.000  &  0.000  &  0.000  &  0.000  \\
  0.025  &  0.105  &  0.106  &  0.109  \\
  0.050  &  0.152  &  0.155  &  0.160  \\
  0.075  &  0.192  &  0.195  &  0.201  \\
  0.100  &  0.227  &  0.230  &  0.238  \\[5pt]
  0.150  &  0.290  &  0.295  &  0.304  \\
  0.200  &  0.350  &  0.354  &  0.365  \\
  0.250  &  0.406  &  0.411  &  0.422  \\
  0.300  &  0.461  &  0.466  &  0.478  \\[5pt]
  0.400  &  0.568  &  0.573  &  0.586  \\
  0.500  &  0.673  &  0.677  &  0.691  \\
\end{tabular}
\label{tab:1}
\end{table}

It is then important that the corresponding results 
obtained through Eq. (\ref{eq:8}), listed in the second column of 
Table \ref{tab:1}, are in agreement with the DMRG calculation.
Since the relative accuracy is expected to further improve at 
nonzero crystal momentum $q$, one must conclude that the magnon 
dispersion (\ref{eq:2}) is sufficiently accurate for all practical 
purposes. Therefore, any disagreement between theory and experiment 
should be attributed to other reasons. In particular, one should note 
in Table \ref{tab:1} that the $1/n$ results slightly underestimate
the DMRG data and hence the latter cannot be used to eliminate 
the remaining small disagreement with the experimental data shown 
in the insets of Fig. \ref{fig:1}.

Next we comment on the relative validity of the 
standard semiclassical theory based on a $1/s$ expansion.
The corresponding harmonic approximation of the magnon 
dispersion is clearly inaccurate, as is apparent in the estimate 
of anisotropy $A=4.5 $ K encountered in the early 
literature.\cite{1}
However, the semiclassical prediction can be significantly 
improved by including the first (anharmonic) $1/s$ correction.
At zero field, a completely analytical calculation is possible
and may be found in Ref. \onlinecite{14}. For nonzero field, the anharmonic
correction is expressed in terms of complicated integrals
that cannot be computed analytically. Therefore, for simplicity,
the main point is made here by considering only the $q=0$ magnon gap
which can be written as
\begin{eqnarray}
\label{eq:9}
&&
G = G_0 \left[	1 + \delta/s + O(1/s^2)  \right],
\nonumber \\
&&
G_0 = s J \left[ h(h+2 \alpha) \right]^{1/2},
\qquad
\delta = \frac{\alpha}{h+2\alpha} \left(
	\frac{1}{2} - I 
\right), 
\\
&& 
I = \frac{1}{\pi} \int_0^{\pi} dq
\frac{ 1 -\cos q +h/2 +\alpha/4 \qquad}
	{ \left[
		(1 -\cos q +h/2)
		(1 -\cos q +h/2 +\alpha)
	\right]^{1/2} },
\nonumber
\end{eqnarray}
where the rationalized field is now defined as 
$h=g_{\perp} \mu_B H/s J$ which differs from the definition given
in Eq. (\ref{eq:3}) by a factor that becomes unimportant for $s=1$.
$G_0$ is the (harmonic) classical approximation 
and $\delta$ provides the first anharmonic correction which amounts 
to about $15 \sim 20 \% $ of the total answer. Numerical values for 
the gap calculated from Eq. (\ref{eq:9}), applied for $s=1$, are listed
in the fourth column of Table \ref{tab:1}.
These values overestimate the DMRG data by a wider margin than the 
{\it harmonic} $1/n$ approximation underestimates the same data.
Therefore, we again conclude that the magnon dispersion (\ref{eq:2})
and the magnon gap (\ref{eq:8}) provide the most accurate description.

Finally, we mention that an $1/n$ expansion is also
possible in the case of a field parallel to the $c$-axis, along
the lines outlined in the Appendix of Ref. \onlinecite{6}. Such a 
possibility will not be pursued further in the present paper,
except for a minor application in Sec. \ref{sec:3:B},
mainly because we do not know of an experimental measurement
of the magnon dispersion for this field orientation.

\section{Thermodynamics}
\label{sec:3}

The most straightforward method for calculating the partition 
function is a complete numerical diagonalization of the Hamiltonian 
on finite chains. The size of the resulting matrices
is $3^N \times 3^N$ and grows exponentially with the total number of sites
$N$. Therefore, a calculation is possible only on short chains while 
a reliable extrapolation to larger values of $N$ is difficult.

More powerful numerical methods proceed with the construction
of a quantum transfer matrix (QTM) obtained by an $M$-step
Trotter decomposition. An explicit calculation was initially performed 
via Quantum Monte Carlo sampling \cite{15} and was also
limited to short chains ($N=16$) and a relatively small number 
of Trotter steps ($M=12$). This procedure led to reasonable results
for the magnetization and susceptibility, but the calculation of 
the specific heat was plagued by large statistical errors.
\begin{figure}
\centerline{\hbox{\psfig{figure=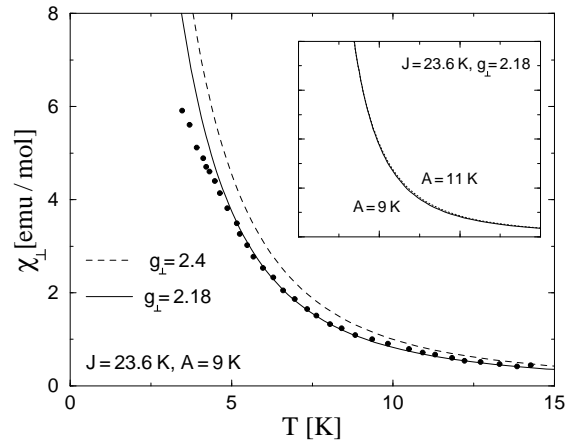,width=7.5cm}}}
\caption{
Comparison of TMRG predictions for the temperature 
dependence of the zero-field transverse susceptibility
$\chi_{\perp}$ with experimental data from Ref. \protect\onlinecite{16} (solid
circles). The dashed line corresponds to the standard set of parameters
of Eq. (\protect\ref{eq:6}) and the solid line to a lower value of the 
gyromagnetic ratio ($g_{\perp}=2.18$). The inset illustrates 
the calculated susceptibility for two values of anisotropy,
$A=9 $ K (solid line) and $A=11 $ K (dotted line), which
lead to virtually identical results.
}
\label{fig:2}
\end{figure}

A more systematic QTM calculation was later accomplished \cite{4} 
on long chains ($N \sim 150$) by limiting the number of Trotter steps
($M \leq 6$) which allows an accurate diagonalization of the matrices
involved in the Trotter decomposition. At first sight, a small $M$
limits the calculation to high temperatures. However, Delica et al. \cite{4}
extrapolate their results for $M=4,5$ and $6$ to higher values of $M$
and thus obtain thermodynamic quantities that are expected 
to be accurate to within a few percent in the temperature region
$T>0.16J \simeq 4 $ K. This restriction is not crucial for application
to $\mathrm{CsNiF_3}$ in view of the 3D-ordering transition
below $T_N =2.7 $ K which limits the validity of the 1D
model anyway.

Our calculation is based on the recently developed 
transfer matrix renormalization group (TMRG)
algorithm \cite{8,9,10} which concentrates on the largest eigenvalue
of the QTM and thus addresses directly the infinite-chain
limit. Furthermore, the number of Trotter steps can
be chosen to be large ($M \sim 160$) if the resulting huge
matrices are diagonalized by a judicious truncation to a
finite number of important states chosen in a manner 
analogous to that employed in the earlier DMRG calculation
of ground-state properties.\cite{7} 
The explicit numerical results discussed in the remainder
of this paper were stabilized to an accuracy better than line thickness,
down to temperature as low as $T=0.02 J \simeq 0.5 $ K which 
is one order of magnitude lower than the lowest temperature
reached in earlier calculations. 
We find that the results of Delica 
et al. \cite{4} are reliable, within the anticipated limits of accuracy,
whereas the more recent elaborate semiclassical calculation of 
Cuccoli et al. \cite{5} is not very accurate over the temperature 
region of current interest.

\subsection{Field perpendicular to $c$}
\label{sec:3:A}

We begin with the discussion of the temperature dependence
of the zero-field transverse susceptibility $\chi_{\perp}$ measured
sometime ago by Dupas and Renard. \cite{16} The TMRG result for the 
standard set of parameters (\ref{eq:6}) is depicted by a dashed line 
in Fig. \ref{fig:2} and is seen to systematically deviate from the 
experimental data. The agreement with experiment for this
set of parameters claimed by Cuccoli et al. \cite{5} is 
due to inaccuracies in their calculation, a point that will be 
made more explicit in our subsequent discussion of the 
specific heat.

Now, the transverse susceptibility $\chi_{\perp}$ is found to be
largely insensitive to the specific strength of anisotropy, as
demonstrated in the inset of Fig. \ref{fig:2}. On the other hand, 
$\chi_{\perp}$ depends quadratically on the gyromagnetic ratio
$g_{\perp}$ and is thus very sensitive to its specific value.
It is then important that a reasonable agreement with the data is achieved
for the same value $g_{\perp}=2.18$ obtained by our spinwave
analysis of Sec. \ref{sec:2}, as shown by the solid line in the main
frame of Fig. \ref{fig:2}. The remaining systematic departure from the data
observed for $T \lesssim 5 $ K could be due to a gradual onset
of 3D ordering at low temperatures.

\begin{figure}
\centerline{\hbox{\psfig{figure=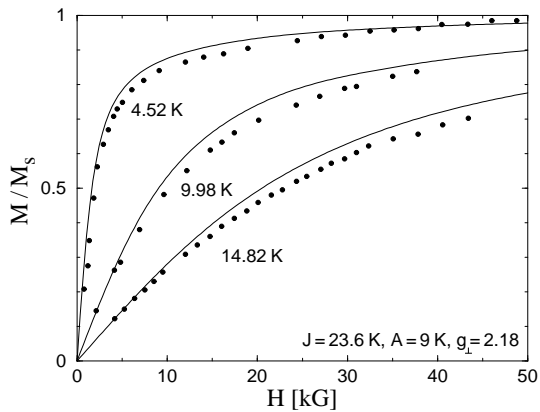,width=7.5cm}}}
\caption{
Comparison of TMRG predictions for the field dependence 
of the magnetization $M$ at selected temperatures with
experimental data from Ref. \protect\onlinecite{4} (solid circles). $M_s$ is
the saturation magnetization, and the specific choice of 
parameters is discussed in the text.
}
\label{fig:3}
\end{figure}

The above choice of the gyromagnetic ratio is further challenged 
by comparing, in Fig. \ref{fig:3}, the TMRG prediction for the 
field dependence of the magnetization with experimental data 
taken at selected temperatures.\cite{4} The specific value of $A$
chosen in Fig. \ref{fig:3} is not important because the transverse
magnetization is also not particularly sensitive to the 
strength of anisotropy. But the relative low value $g_{\perp}=2.18$
was again important to improve agreement with the data.
Yet a significant disagreement between theory and experiment
is apparent in Fig. \ref{fig:3}, even at relatively high temperatures.
The lower value $g_{\perp}=2.1$ employed in Ref. \onlinecite{4} reduces
but does not eliminate the discrepancy.
An attempt to remedy this situation by incorporating a phenomenological
interchain interaction leads to a deterioration 
of the corresponding theoretical prediction for the zero-field
transverse susceptibility.\cite{4}

\begin{figure}
\centerline{\hbox{\psfig{figure=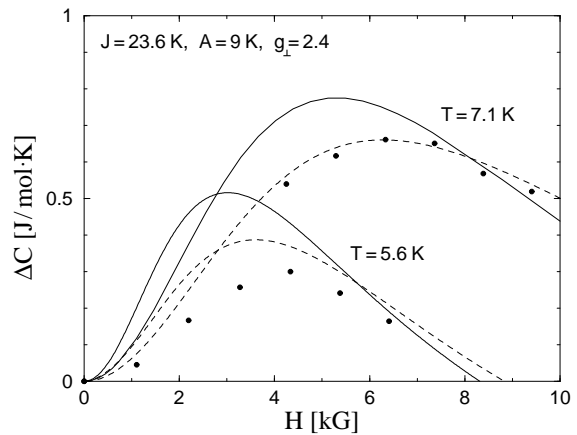,width=7.5cm}}}
\caption{
Comparison of TMRG predictions for the excess specific 
heat (solid lines) with experimental data from Ref. \protect\onlinecite{17}
(solid circles) for two typical values of temperature. The 
dashed lines depict the corresponding theoretical results of
Ref. \protect\onlinecite{5} for the same set of parameters given by Eq. 
(\protect\ref{eq:6}).
}
\label{fig:4}
\end{figure}
We next discuss the specific heat $C=C(T,H)$ which was
measured experimentally by Ramirez and Wolf.\cite{17} In fact,
most of the attention was concentrated on the {\it excess} specific
heat
\begin{equation}
\label{eq:10}
\Delta C = C(T,H) -C(T,0)
\end{equation}
viewed as a function of field $H$ at some specified temperature
$T$. An elementary argument based on the dilute-magnon
approximation suggests that $\Delta C$ is negative and decreases
with increasing field, because the magnon dispersion discussed
in Sec. \ref{sec:2} increases monotonically with $H$ for all values of
the crystal momentum $q$. Nevertheless, the experiment revealed
that $\Delta C$ rises to a positive maximum at some field $H_{max} \sim T^2$
before it begins to decrease and eventually reach negative
values for stronger fields. A possible explanation of this unexpected
behavior could be that the dilute-magnon approximation
breaks down in the actual temperature range of
the experiment, or ``nonlinear modes'' are activated
in addition to magnons. Whence the beginning of a
long debate concerning the possible relevance of sine-Gordon
kinks, at least in some approximate sense.\cite{4,5}

One of the advantages of an accurate numerical algorithm
such as TMRG is that potential nonlinear effects are automatically
taken into account. Our results for the excess specific
heat calculated for the standard choice of parameters 
given in Eq. (\ref{eq:6}) are depicted in Fig. \ref{fig:4} 
for two characteristic
values of temperature actually employed in the experiment.\cite{17}
Inspite of the overall qualitative agreement, significant quantitative
differences are apparent in Fig. \ref{fig:4} for both values of
the temperature. We were thus surprised to note that the theoretical
results of Cuccoli et al. \cite{5,18} for the same set of parameters,
depicted by dashed lines in Fig. \ref{fig:4}, are in agreement
with the data for the specific temperature $T=7.1 $ K. On the 
other hand, our results agree with those given by Delica et al. 
\cite{4} for the same set of parameters, except for some 
minor (a few percent) differences anticipated by the introductory
remarks of this Section. As mentioned already, a 
similar criticism applies to the calculation of the transverse
susceptibility by Cuccoli et al.\cite{5}. We must thus conclude
that the semiclassical nature of their method does not 
allow a completely accurate calculation in this temperature
range and the claimed agreement with experiment 
is fortuitous.

\begin{figure}
\centerline{\hbox{\psfig{figure=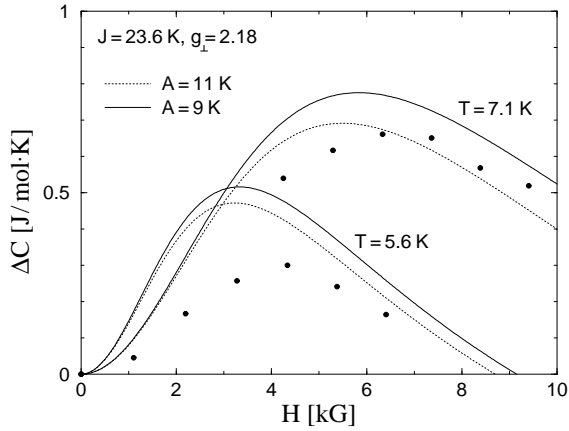,width=7.5cm}}}
\caption{
Comparison of TMRG predictions for the excess specific 
heat, for two different sets of parameters, with experimental 
data from Ref. \protect\onlinecite{17} (solid circles).
}
\label{fig:5}
\end{figure}

It is now interesting to examine whether or not the 
alternative set of parameters given in Eq. (\ref{eq:7}) may be used
to eliminate the observed differences. In fact, our results
quoted in Fig. \ref{fig:5}, together with those given in Fig. \ref{fig:4}
of Ref. \onlinecite{4} for yet another set of parameters, suggest that 
an accurate fit of the data is not possible for any
reasonable choice of parameters.

Nevertheless, the main qualitative features of the experimental 
data are reproduced by the theoretical calculation.
Therefore, it is important to examine further within the 1D
model the mechanism by which the simple spinwave argument
given earlier in the text is reconciled with a positive excess
specific heat. We first consider the quantity
\begin{equation}
\label{eq:11}
-T \ln (T^{3/2} C) = G +G_1 T +G_2 T^2 + \ldots ,
\end{equation}
where the expansion in the right-hand side presumes that the 
low-temperature thermodynamics is dominated by magnons with a $q=0$
energy gap equal to $G$. A detailed TMRG calculation of 
the left-hand side of Eq. (\ref{eq:11}) for low temperatures down to $T=0.02J$
reveals a behavior that is indeed consistent with the right-hand 
side of the same equation. Putting it in more practical
terms, an extrapolation to $T=0$ using a second-degree
polynomial to fit the low-temperature numerical data yields estimates
of the magnon gap $G$ which are in agreement with the direct
DMRG calculation given in Table \ref{tab:1}. A curious fact is that 
the present calculation gives values for the gap that are even 
closer to the $1/n$ results of Table \ref{tab:1}, but this may be an 
artifact of the specific second-order interpolation scheme.

The implied normal spinwave behavior of this easy-plane
ferromagnetic chain should be contrasted with the 
low-temperature anomalies discovered by Johnson and Bonner \cite{19}
in an easy-axis ferromagnetic chain and recently confirmed
by a TMRG calculation.\cite{20} The absence of such anomalies 
in the present model reinforces the need for explaining the excess
specific heat in simple terms.

In the remainder of this subsection we find it convenient
to work exclusively with the rationalized parameters $\alpha$ and $h$
of Eq. (\ref{eq:3}) whereas the temperature $\tau= T/J$ is measured 
in units of the exchange constant $J$. The corresponding
absolute specific heat per lattice site is denoted 
by $c=c(\tau,h)$ and the excess specific heat by 
$\delta c=c(\tau,h)-c(\tau,0)$.

\begin{figure}
\centerline{\hbox{\psfig{figure=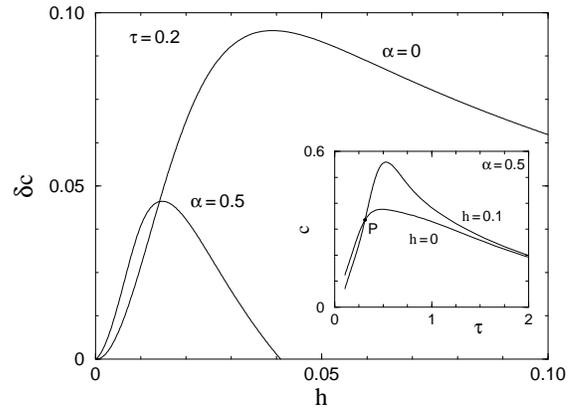,width=7.5cm}}}
\caption{
TMRG calculation of the excess specific heat $\delta c$ for a typical
anisotropy ($\alpha=0.5$) and for the isotropic spin-$1$ ferromagnetic
chain ($\alpha=0$). The inset depicts the temperature dependence
of the absolute specific heat $c$ for two field values, $h=0$
and $0.1$, and anisotropy $\alpha=0.5$. All quantities shown in
this figure are  expressed in rationalized units.
}
\label{fig:6}
\end{figure}
The inset of Fig. \ref{fig:6} illustrates the calculated temperature
dependence of the specific heat $c$ for a typical anisotropy
$\alpha=0.5$ and two field values; $h=0$ and $0.1$. It is clear 
that a nonzero field causes a depression of the specific 
heat at low temperatures thanks to the opening of a 
finite magnon gap. This is the expected normal spinwave 
behavior, as predicted by the usual dilute-magnon approximation.
What is not accounted for by dilute magnons
is the crossing of the $h=0$ and $h=0.1$ curves at a point $P$
that corresponds to a specific temperature $\tau$ which 
depends on $h$. In particular, $P$ is located near the 
origin for small $h$ and moves outward with increasing $h$.
This crossing is precisely the origin of the positive excess
specific heat at low $h$, as demonstrated again by the $\alpha=0.5$
solid curve in the main frame of Fig. \ref{fig:6} for the specific 
temperature $\tau=\tau_0=0.2$.

Indeed, for any fixed $\tau_0$, the crossing point $P$
occurs at some $\tau<\tau_0$ for sufficiently weak fields, and 
thus leads to positive $\delta c$ at $\tau=\tau_0$. With increasing field
the point $P$ moves to the right and the corresponding 
temperature $\tau$ eventually overtakes $\tau_0$, thus leading to
negative $\delta c$ at $\tau=\tau_0$ for sufficiently strong fields. The 
described picture is valid for any choice of $\tau_0$, and is confirmed
by all of our numerical experiments. Therefore,
the explanation of a positive $\delta c$ at low fields is equivalent
to ascertaining the robust enhancement of the absolute
specific heat $c$ with increasing field, inspite of its initial 
depression by the field dependent magnon gap.

At this point one could invoke the popular sine-Gordon
approximation to argue that the crossing mechanism described
in the preceding paragraph is due to the activation of kinks
or other nonlinear modes in addition to magnons. We think
that such an interpretation is dubious simply because the 
same mechanism occurs also in the {\it isotropic} Heisenberg
chain, as illustrated by the $\alpha=0$ line in Fig. \ref{fig:6}.
In fact, the effect is strongly pronounced in the isotropic limit,
even though a sine-Gordon approximation is clearly out of question.

Therefore, we return to the described crossing mechanism
and attempt to explain it by more elementary means.\cite{21} The
absolute specific heat satisfies the obvious identity
\begin{equation}
\label{eq:12}
\int_0^{\infty} d \tau \, c(\tau,h) = u(\infty,h) -u(0,h),
\end{equation}
where $u(\tau,h)$ is the internal energy at temperature $\tau$ and field $h$.
A corresponding identity for the excess specific heat is obtained
by applying Eq. (\ref{eq:12}) twice:
\begin{equation}
\label{eq:13}
\int_0^{\infty} d \tau \, \delta c(\tau,h) = 
\left[ u(\infty,h) -u(\infty,0) \right]
+ \left[ u(0,0) -u(0,h) \right].
\end{equation}
A significant simplification occurs in the limit of an 
isotropic ferromagnetic chain for which the field dependence of the 
energy levels is simply a linear Zeeman shift $m h$, with
$m=0, \pm 1, \pm 2, \ldots$. Therefore the field dependence averages out
of the infinite-temperature internal energy $u(\infty,h)$, which
is the sum of all energy levels, and $u(\infty,h) -u(\infty,0)=0$.
If we further recall that $e(h)=u(0,h)$ is the ground-state
energy at field $h$, we obtain the elementary sum rule
\begin{equation}
\label{eq:14}
\int_0^{\infty} d \tau \, \delta c(\tau,h) = e(0) -e(h) = h,
\end{equation}
where we have also invoked the known energy of 
the fully polarized ferromagnetic ground state.

The obvious consequence of Eq. (\ref{eq:14}) is that positive values 
of $\delta c$ are the rule rather than the exception. In particular,
the initial depression of the specific heat ($\delta c<0$) at low 
temperatures, due to the opening of a magnon gap at
finite field, is overwhelmed by positive values of
$\delta c$ attained at higher temperatures also thanks to the 
applied field. This explains the gross features of the 
crossing mechanism described earlier in the text and 
concludes our discussion of the excess specific heat.

\subsection{Field parallel to $c$}
\label{sec:3:B}

The case of a field parallel to the $c$-axis is equally interesting
but the corresponding experimental work has not been
as extensive. We begin with the discussion of the temperature 
dependence of the zero-field longitudinal susceptibility. A
notable feature of $\chi_{\parallel}(T)$ is that it must approach a finite 
value in the limit $T \rightarrow 0$. A simple estimate of this value 
is obtained by a straightforward classical argument. In the 
presence of a field ${\bf H}=(0,0,H)$ the classical ground state 
is such that all spins form an angle $\theta$ with the $c$-axis
calculated from $\cos \theta = g_{\parallel} \mu_B H/2A$. Therefore, the $T=0$
magnetization is given by $M=N g_{\parallel} \mu_B \cos \theta$ and the 
susceptibility by
\begin{equation}
\label{eq:15}
\chi_{\parallel}^{cl}(T=0) = \frac{1}{2A} (N g_{\parallel}^2 \mu_B^2),
\end{equation}
where $N$ is the total magnetic sites and $g_{\parallel}$ is the 
gyromagnetic ratio for a field applied along the $c$-axis.
\begin{figure}
\centerline{\hbox{\psfig{figure=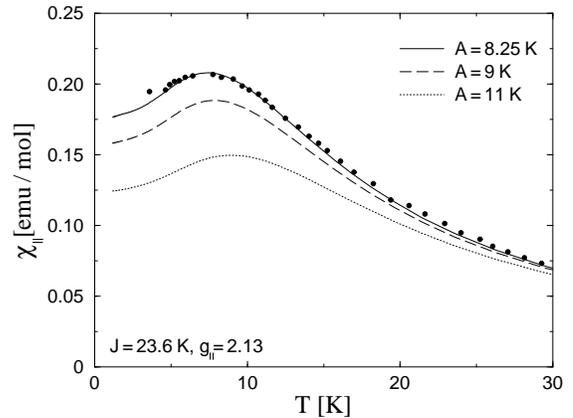,width=7.5cm}}}
\caption{
Comparison of TMRG predictions for the temperature 
dependence of the zero-field longitudinal susceptibility $\chi_{\parallel}$
with experimental data from Ref. \protect\onlinecite{16} (solid circles).
}
\label{fig:7}
\end{figure}

Of course, numerical estimates based on the above classical
result are not expected to be accurate, for reasons similar
to those explained in Sec. \ref{sec:2}. However, a more accurate
prediction may again be obtained through the $1/n$ expansion. To 
leading order, the $T=0$ magnetization is calculated as the 
expected value of the azimuthal spin in the Hartree variational
ground state given in the Appendix of Ref. \onlinecite{6}. Restricting 
that calculation to weak fields one may extract the $T=0$
longitudinal susceptibility
\begin{equation}
\label{eq:16}
\chi_{\parallel}^{1/n}(T=0) = 
\frac{1}{A} 
\left( 1 -\frac{A}{4J} \right)
(N g_{\parallel}^2 \mu_B^2).
\end{equation}
The main difference from Eq. (\ref{eq:15}) is an overall factor of $2$,
which is essentially the same factor that caused the low
estimate $A=4.5 $ K in the early literature,\cite{1} 
in addition to some mild dependence on the exchange constant.
In any case, the main conclusion is that $\chi_{\parallel}$ is more
sensitive to the value of the anisotropy constant 
$A$ than to the exchange constant $J$, a situation that is 
reverse to the one encountered in Sec. \ref{sec:3:A}.

\begin{figure}
\centerline{\hbox{\psfig{figure=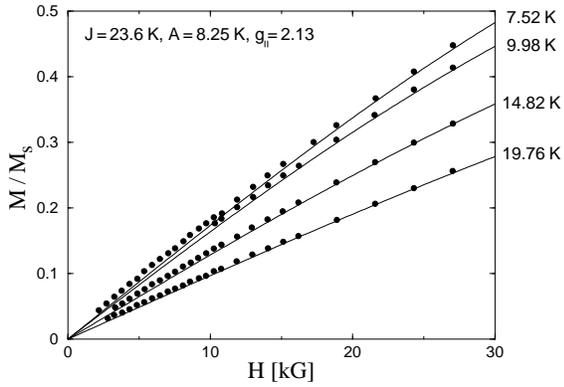,width=7.5cm}}}
\caption{
Comparison of TMRG predictions for the field dependence of
the magnetization $M$ at selected temperatures with experimental
data from Ref. \protect\onlinecite{4} (solid circles). The field is applied
along the $c$-axis and $M_s$ is the saturation magnetization.
}
\label{fig:8}
\end{figure}

Therefore, the longitudinal susceptibility is an ideal
physical quantity to yield a sensible estimate of the 
anisotropy constant $A$, provided that an accurate value
for $g_{\parallel}$ is also available. The latter is fixed here by appealing to
a theoretical estimate \cite{16} of the difference $g_{\perp}-g_{\parallel} \simeq 5\times 10^{-2}$ which
leads to $g_{\parallel}=2.13$ if we adopt our earlier value for the transverse 
gyromagnetic ratio $g_{\perp}=2.18$. The corresponding TMRG 
calculation of  $\chi_{\parallel}(T)$ is illustrated in Fig. \ref{fig:7} for various reasonable
choices of $A$. The experimental data \cite{16} are well reproduced for the 
set of parameters
\begin{equation}
\label{eq:17}
J=23.6 \ \mathrm{K},
\qquad
A=8.25 \ \mathrm{K},
\qquad
g_{\parallel}=2.13,
\end{equation}
which is closer to the set employed by Delica et al. \cite{4}.
In addition, the field dependence of the magnetization measured at selected
temperatures \cite{4} agrees with our TMRG calculation
without further fit of parameters, as demonstrated in Fig. \ref{fig:8}.

Incidentally, for this choice the classical result (\ref{eq:15}) yields
$0.10 $ emu/mol and the leading $1/n$ approximation (\ref{eq:16})
gives $0.19 $ emu/mol. These values should be compared
with $\chi_{\parallel}(T=0) \simeq 0.175 $ emu/mol extracted by a visual extrapolation
of the solid curve in Fig. \ref{fig:7} to $T=0$. 
Including the $1/n$ correction produced by zero-point fluctuations
in Eq. (\ref{eq:16}) will bring its prediction to the same level of accuracy
with the magnon gap discussed in Table \ref{tab:1}.

It is now interesting to take this calculation into the region
of strong fields where the ground state becomes completely ordered
along the $c$-axis. Such a ferromagnetic state is actually an
exact eigenstate of the Hamiltonian for any strength of the field
$H$. But the corresponding magnon gap
\begin{equation}
\label{eq:18}
G = g_{\parallel} \mu_B H -A
\end{equation}
is positive only for $H>H_c$ where
\begin{equation}
\label{eq:19}
H_c = A/g_{\parallel} \mu_B
\end{equation}
is the critical field beyond which the fully ordered state is the 
absolute ground state. The gap vanishes for all $H<H_c$ because
the corresponding magnon is a Goldstone mode associated
with the axial symmetry for this field orientation.

\begin{figure}
\centerline{\hbox{\psfig{figure=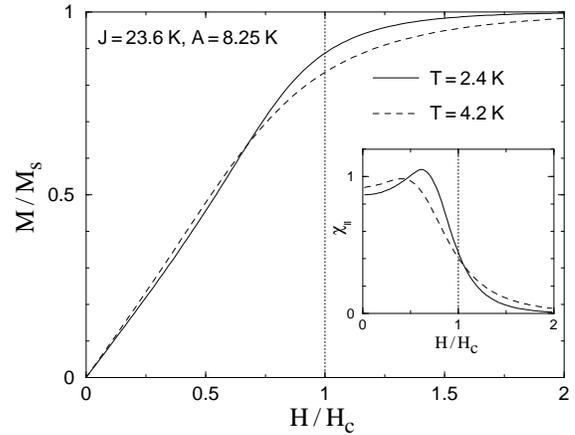,width=7.5cm}}}
\caption{
TMRG calculation of the field dependence of the magnetization
$M$ for a wide field range and two typical values of temperature.
The inset displays the corresponding results
for the field dependence of the susceptibility. The field
is applied along the $c$-axis and 
the critical field $H_c$ is estimated to be $58 $ kG for $g_{\parallel}=2.13$.
}
\label{fig:9}
\end{figure}

For the set of parameters (\ref{eq:17}) one finds that $H_c=58 $ kG,
in reasonable agreement with the value $62.5 $ kG estimated
from an experiment of A. Miedan which is quoted in Ref. \onlinecite{16}
but is apparently unpublished. According to the description of
Dupas and Renard,\cite{16} Miedan measured the field dependence 
of the magnetization at $T=4.2 $ K and extracted $H_c$ from the
observed bending of the $M(H)$ curve. Although we do not know
the details of this experiment, we have calculated the $M(H)$
curve at $T=4.2 $ K for a wide field range and the result 
is depicted by a dashed line in Fig. \ref{fig:9}. Interestingly, the 
bending of the $M(H)$ curve
is not predicted to be especially sharp at this temperature,
as is apparent in the corresponding susceptibility displayed
also by a dashed line in the inset of Fig. \ref{fig:9}. In other words,
if the location of the maximum of the susceptibility were
taken as an estimate of the critical field $H_c$, the latter
would have been severely underestimated. The situation
improves slowly at lower temperatures, as indicated by
the solid lines in Fig. \ref{fig:9} which correspond to $T=2.4 $ K;
i.e., to a temperature that is already below the 
3D-transition temperature $T_N=2.7 $ K.

It is clear that we cannot go farther with our theoretical
arguments without explicit knowledge of detailed 
experimental data on $M(H)$ in this field region. We thus 
conclude the discussion of magnetization with a comment
concerning an apparent contradiction between the results
of Fig. \ref{fig:9} and those given earlier in Fig. \ref{fig:8} for lower
field strengths. Indeed, Fig. \ref{fig:8} suggests that the magnetization
$M(H)$ for any given field $H$ decreases with increasing temperature,
as expected, while Fig. \ref{fig:9} indicates that a relative
crossing occurs between any two $M(H)$ curves. The resolution
of this apparent paradox lies in the fact that the values of temperature
employed in Fig. \ref{fig:8} are all greater than the temperature
$T \simeq 7.5 $ K, at which the maximum of the zero-field
susceptibility of Fig. \ref{fig:7} occurs, while those of Fig. 
\ref{fig:9} are smaller.

\begin{figure}
\centerline{\hbox{\psfig{figure=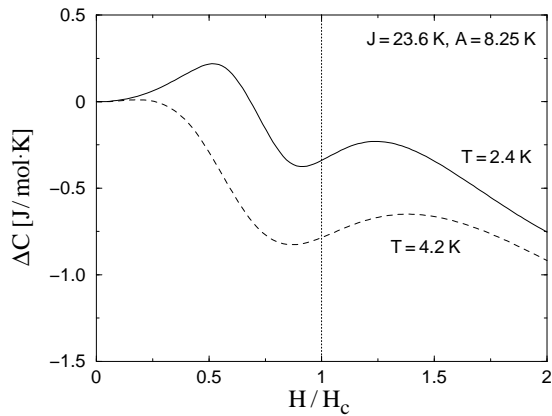,width=7.5cm}}}
\caption{
TMRG calculation of the excess specific heat for a 
wide field range and two typical values of temperature.
The field is applied along the $c$-axis and the 
critical field $H_c$ is estimated to be $58 $ kG for 
$g_{\parallel} =2.13$.
}
\label{fig:10}
\end{figure}

Finally, we discuss the specific heat in a field parallel 
to the $c$-axis. It appears that no measurements have been
made for this field orientation but could prove to be feasible 
in the future.\cite{22} Our TMRG calculation of the excess 
specific heat is illustrated in Fig. \ref{fig:10} for the two values of 
temperature employed in our preceding discussion of the
magnetization. The characteristic double peak near the 
critical field $H_c$ was anticipated by earlier work \cite{21}
based on a classical transfer matrix calculation, on the 
known exact solution for a spin-$\frac{1}{2}$ XY chain, as well as 
on an accurate numerical solution for a spin-$\frac{1}{2}$ XXZ
chain based on the Bethe Ansatz. The calculated double 
peak is also a clear departure from the corresponding prediction
of the dilute-magnon approximation \cite{21} and could eventually
be observed in $\mathrm{CsNiF_3}$. An unfortunate feature of Fig. \ref{fig:10}
is that a strongly pronounced double peak is predicted to 
occur in the low-temperature region where the 1D model is no longer applicable.

\section{Conclusion}
\label{sec:4}

We have presented a more or less complete calculation
of the dynamics and the thermodynamics associated
with the spin-$1$ Hamiltonian (\ref{eq:1}). The $T=0$ dynamics
is efficiently described by an $1/n$ expansion
whose full potential has not yet been explored. For example,
an accurate calculation of the magnon dispersion for a field 
parallel to the $c$-axis is also possible but has not been carried 
out mainly because there seems to have been no 
experimental effort in that direction.

On the other hand, the thermodynamics is calculated by 
a powerful TMRG method which has opened the way
to obtain accurate theoretical predictions for a wide class
of quantum magnetic chains. Suffice it to say that our 
present algorithm may be trivially adjusted to handle 
spin-$1$ Haldane-gap antiferromagnets in the 
presence of anisotropy and external fields. Even 
in the case of completely integrable spin-$\frac{1}{2}$ chains, for which
the Bethe Ansatz applies, the calculation of the thermodynamics 
is far from trivial.\cite{23} Nevertheless, TMRG can be applied 
in a straightforward manner irrespective of complete
integrability.\cite{20}

The extent to which the 1D Hamiltonian (\ref{eq:1}) may 
describe the magnetic properties of $\mathrm{CsNiF_3}$ has been debated 
on several occasions. Our calculations confirm the general
conclusion that the 1D model accounts for the main 
features of all available experimental data. But it 
is also clear that departures from ideal model behavior
are present, especially at low temperatures
approaching the 3D-ordering transition temperature $T_N=2.7 $ K.

\acknowledgements
XW and XZ acknowledge financial support by the Swiss National Fund, the 
University of Fribourg, and the University of Neuch\^atel.

\end{document}